\input amstex

\documentstyle{amsppt} \magnification1200 \TagsOnRight
\NoBlackBoxes
\pagewidth{30pc} \pageheight{47pc} 

\input epsf.tex

\def\today{9 April 1997}

\def\SU{\operatorname {SU}}
\def\SO{\operatorname {SO}}

\def\Z{{\Bbb Z}}
\def\R{{\Bbb R}}
\def\identity{{\Bbb I}}
\def\d{{\text{d}}}

\topmatter
\title  An algebraic interpretation of the Wheeler-DeWitt equation\endtitle
\author John W. Barrett\\ Louis Crane\endauthor
\date\today\enddate
\address
Department of Mathematics,
University of Nottingham,
University Park,
Nottingham,
NG7 2RD, UK
\endaddress
\email jwb\@maths.nott.ac.uk \endemail
\address
 Mathematics Department,
Kansas State University,
Manhattan KS, 66502, USA
\endaddress
\email crane\@math.ksu.edu\endemail

\abstract
We make a direct connection between the construction of three
dimensional topological state sums from tensor categories and three
dimensional quantum gravity by noting that the discrete version of the
Wheeler-DeWitt equation is exactly the pentagon for the associator of
the tensor category, the Biedenharn-Elliott identity. A crucial role is
played by an asymptotic formula relating 6j-symbols to rotation matrices given
by Edmonds.
\endabstract

\thanks This research was supported by an EPSRC visiting fellowship.\endthanks
 \endtopmatter

\document

\head Introduction \endhead

It has been known since the work of Ponzano and Regge \cite{1} that the
quantum theory of gravity in three Euclidean dimensions can be
described by means of the evaluation of spin networks, which are
combinations of representations of the Lie group $\SU(2)$ under the
tensor product in the category of representations of $\SU(2)$. The ideas were
elaborated by Hasslacher and Perry \cite{2}.

The evaluation of a spin network on a sphere can
be calculated by triangulating the three-dimensional ball bounded by the
diagram and
labelling all interior edges with representations. Ponzano and Regge gave a
formula as a
sum over all such labellings of the product of the 6J symbols
associated to the tetrahedra of the triangulation. The edge labels $j$ are
identified with half-integers.

$$\multline\Psi(\text{boundary edge labels})=\\
\sum_{\text{interior edge labels}}
(-1)^\chi\left(\prod_{\text{tetrahedra}}\text{6j-symbol}\right)
\prod_{\text{interior edges}}(2j+1)\endmultline\tag1$$
where the integer $\chi$ is a function of the edge labels. If the sum is
finite, then it is independent of the triangulation in the interior of the
ball. A more detailed account of this has appeared in \cite{3}. The formula can
also be applied to obtain a wavefunction for other 3-manifolds more general
than the ball.

Subsequent developments have put this formula into a broader context.
Turaev and Viro \cite{4} showed that the analogous formula, using the
truncated category of representations of a quantum group at a root of
unity gives rise to a three dimensional topological quantum field
theory (TQFT), closely related to the Chern Simons theory of Witten \cite{5}.

Barrett and Westbury \cite{6,7}, have demonstrated that the natural
generalisation of this construction is a tensor category with unit and
duals and only a very weak and natural assumption on the left and
right traces. In particular the braiding or symmetry of the tensor
product is unnecessary. The useful generalization of 6J symbols is the
associator isomorphism of the tensor category.

Thus, we have what seems to be a rather mysterious
connection between a very abstract branch of mathematics, namely the
theory of tensor categories, and the very deep physical problem of
quantising general relativity.

The purpose of this paper is to make this connection more direct, and
possibly somewhat clearer. What we show is that the Wheeler-DeWitt
equation for 3d general relativity reduces directly to the pentagon
relation for the tensor category we are using, in the presence of an
asymptotic formula for thin tetrahedra. Since the pentagon \cite{8} is just the
coherence condition for the associator of the tensor product of the
category, we are identifying the laws of motion of general relativity
directly with the most fundamental part of the theory of tensor categories.

This allows us to focus our efforts to make physical interpretations
of theories based on other tensor categories on the discovery of the
thin tetrahedra asymptotic forms of the associator isomorphisms.

This paper should be thought of as shedding some light on the
algebraic program for quantising general relativity outlined in \cite{9}.
The great desideratum, of course, would be to find an analog of the
categorical formula which worked in four dimensions. We have not yet
worked out the four dimensional argument analogous to the one given in
this paper but we believe it should be possible to do so, and that it
might help to recognize whether an algebraic construction of a four
dimensional state sum is related to general relativity.

The Ponzano-Regge wavefunction is defined for triangulations of a 3-manifold
with boundary where the state-sum \thetag1 is a finite sum. The wavefunction is
then the Turaev-Viro partition function with the parameter $q$ set to 1, up to
a normalisation. This shows that when the Ponzano-Regge wavefunction is defined
it depends only on the boundary data and the topology of the 3-manifold, being
independent of the particular triangulation. However the Ponzano-Regge formula
does not define a TQFT because for an arbitrary triangulation the formula
\thetag1 may be infinite.

In any TQFT defined by a state sum in a similar manner to the Turaev-Viro state
sum, there is a vector space $V$ determined by a triangulation of a surface.
The partition functions, or wavefunctions, all lie in a particular subspace
$S\subset V$. While the space $V$ depends on the triangulation, the subspace
$S$ is independent of the triangulation.

One can see on general grounds that $S$ is determined by a set of projectors
$\pi_v$, one associated to each vertex $v$ in the surface, and that $S$ is
determined by the equations
$$\pi_v\psi=\psi\tag 2$$
being simultaneously satisfied. The equations are obviously candidates for the
analogues in this theory of the Hamiltonian constraint equations which are
often supposed to characterise diffeomorphism invariant quantum field theories.

In this paper, we present constraint equations in the form of \thetag2 for the
Ponzano-Regge wavefunction. However the argument is independent of the general
formalism of TQFT. We give an argument that the equations are a quantisation of
the appropriate discrete version of the Wheeler-DeWitt equations, thus
identifying the Ponzano-Regge wavefunction as belonging to a quantisation of a
specific action, namely Einstein gravity.

\head The constraint equations\endhead

The Einstein equation for a metric in three dimensions is that it is locally
flat. On the boundary surface $\Sigma$ of the manifold, this implies that there
are constraints between the extrinsic curvature $K_{ab}$ and the metric
$g_{ab}$ of $\Sigma$. Assuming that the three-dimensional metric is positive
definite, these are
$$ \align\nabla_a{K^a}_b-\nabla_bK&=0\\
\intertext{and} K^2-K^{ab}K_{ab}-R&=0,\tag3\endalign$$
where $K=K^{ab}g_{ab}$, $\nabla$ is the connection for $g$, and $R$ its scalar
curvature. The conventions are those of \cite{10}.

The Einstein action can be used to cast these equations in the framework of
Hamiltonian mechanics \cite{11}, using the momentum density
$$\pi^{ab}=\left(Kg^{ab}-K^{ab}\right)\Omega  \tag4$$
where $\Omega$ is the metric volume measure. Then the constraints are
$$\left(\pi^2-\pi^{ab}\pi_{ab}\right)\Omega^{-1}-R\Omega=0\tag 5$$
and
$$\nabla_a{\pi^a}_b=0.\tag 6$$
For each tangent $h_{ab}$ to the space of metrics on $\Sigma$, canonical
quantisation associates to the momentum
$$\phi=\int_\Sigma\pi^{ab}h_{ab}\tag7$$
the first order differential operator corresponding to $h_{ab}$.

Making this substitution in the Hamiltonians determined by \thetag5 and
\thetag6 does not generally make sense in this infinite-dimensional
configuration space. However this system can be quantised by changing variables
and fixing the gauge in a particular way \cite{12}.

Alternatively, one can make progress by replacing the space of metrics on
$\Sigma$ by a finite-dimensional approximating space. Then canonical
quantisation gives partial differential equations, with some ambiguity due to
the different possible orderings of the operators.

Our quantisation procedure is to use the finite-dimensional space of metrics
determined by Regge calculus \cite{13} and regard quantisation as giving an
approximation to the equations satisfied by the wavefunction, which is already
defined. Thus we are not using quantisation to define a quantum theory, rather
to compare a quantum theory with a classical mechanical one.

The momentum $\phi$ survives the discretisation procedure, and for a metric
fluctation given by changing one particular edge length $l$, $\phi$ is just the
parameter for the extrinsic curvature at the edge discussed by Hartle and
Sorkin \cite{14}. This identity is specific to three dimensions, the momentum
and the extrinsic curvature being logically distinct quantities. If $\Sigma$ is
the boundary of a 3-manifold $M$ with a Euclidean signature metric, this is the
angle of rotation of the outward unit normal from the triangle at one side of
an edge to the normal to the triangle at the other side. It is positive if the
edge is convex and negative if the edge is concave.

\midinsert\centerline{\epsfbox{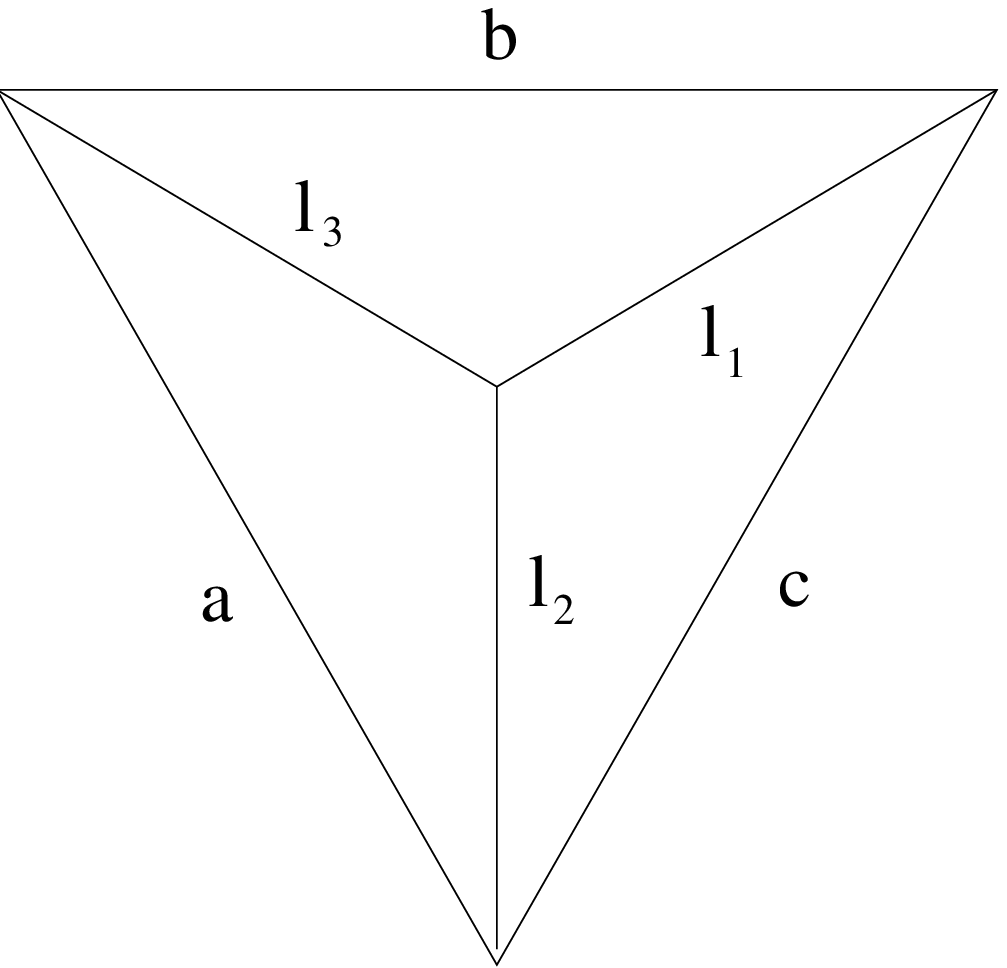}}\botcaption {Figure 1. Part of
$\Sigma$}\endcaption \endinsert

Consider a vertex in $\Sigma$ which belongs to exactly three triangles, such as
appears on the boundary of a tetrahedron (figure 1). The connection in $M$
gives a map $TM_p\to TM_{p'}$ if a path from $p$ to $p'$ in $\Sigma$ is given
which avoids vertices. If an orthonormal frame is chosen at $p$ and $p'$, then
this map determines a rotation in $\SO(3)$
$$\CD
\R^3 @>\text{rotation}>>\R^3\\
@V\text{frame at $p$}VV @VV\text{frame at $p'$}V\\
TM_p @>>>TM_{p'}
\endCD$$

Consider a path which circulates the vertex via six points at which canonical
frames are chosen, as shown in figure 2. At each point, vector $z$ is chosen to
lie along the adjacent edge tangent to $\Sigma$ and $x$ is chosen rotated by a
local orientation of the surface. The frame is completed by vector $y$ which is
the outward normal to $\Sigma$. The six rotation matrices are $R_{12}$,
$R_{23}$, $R_{31}$, which are rotations through $\theta_{12}$, $\theta_{23}$,
$\theta_{31}$ about the y-axis, and $K_1$, $K_2$, $K_3$, which are rotations by
$\phi_1$, $\phi_2$, $\phi_3$ about the z-axis. The $R$'s can be described as
the rotations which take one edge into the next and the $K$'s are the rotations
which make one face parallel to the next.

As for the smooth case, the Einstein equations in three dimensions are that the
metric is flat.
The condition that $M$ is flat is that the holonomy in $M$ is the identity,
$$h=R_{23}K_3R_{31}K_1R_{12}K_2=\identity\in\SO(3).\tag8$$

\midinsert\centerline{\epsfbox{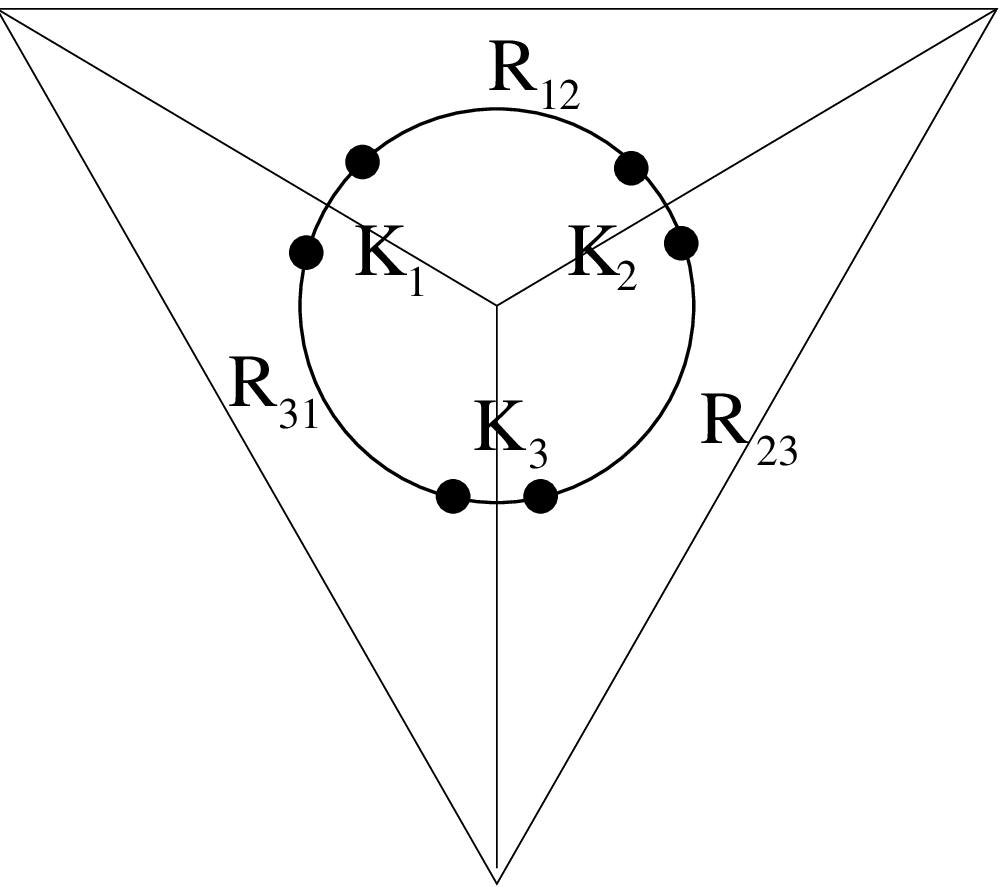}\epsfbox{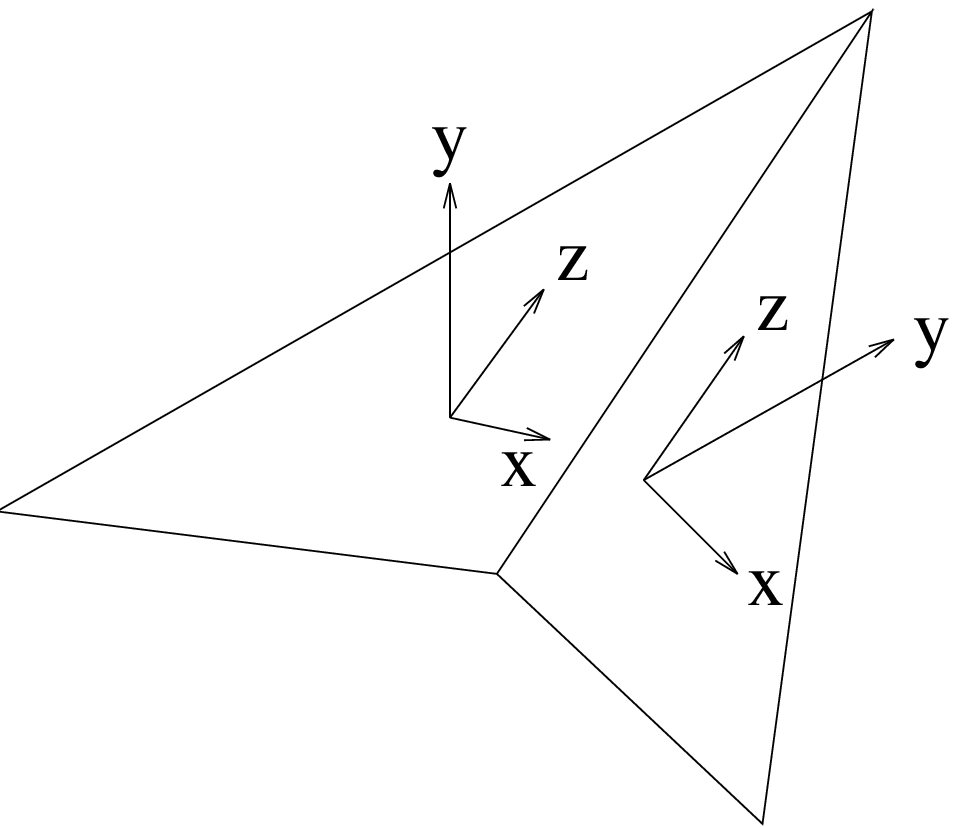}}\botcaption {Figure 2.
Path in $\Sigma$ and choice of frames}\endcaption \endinsert

This gives three equations per vertex of $\Sigma$ which constrain the metric of
$\Sigma$ and its extrinsic curvature. These are the key equations which we wish
to quantise, to give our discrete version of the Wheeler-DeWitt equation.

For the rest of this section there is an argument to demonstrate that the
equations \thetag8 are the discrete analogues of the Hamiltonian constraint
equations. However the development of the rest of the paper is independent of
this argument.

This can be seen by considering small perturbations of the geometry around a
flat configuration $\phi_1=\phi_2=\phi_3=0$. One gets three equations,
corresponding to the three standard basis elements $X$, $Y$, $Z$ of the Lie
algebra of $\SO(3)$, which generate rotations about the $x$, $y$ and $z$ axes.
At the flat configuration, the two-dimensional holonomy is the identity,
$$R_{23}R_{31}R_{12}=\identity\in\SO(3),$$
and a small perturbation results in a small deficit angle $\delta$.
Each edge has associated a Lie algebra element
$$\align L_1&=\bigl(\cos\theta_{12}\bigr)Z+\bigl(\sin\theta_{12}\bigr)X\\
L_2&=Z\\
L_3&=\bigl(\cos\theta_{23}\bigr)Z-\bigl(\sin\theta_{23}\bigr)X\endalign$$
Expanding \thetag8 for small $\phi_1$, $\phi_2$, $\phi_3$, and keeping only the
lowest-order non-zero terms, one gets
$$\phi_1L_1+\phi_2L_2+\phi_3L_3=0$$
and
$$\delta={1\over2}\phi_1\phi_2\sin\theta_{12}.$$
The first of these is two equations, linear in the momenta, given by the
coefficients of $X$ and $Z$. It is a divergence equation, and corresponds to
\thetag6. The second equation, quadratic in the momenta, corresponds to
\thetag5, and is symmetrical in the three edges by virtue of the first
equation, giving the area of a triangle with edges $\phi_1$, $\phi_2$,
$\phi_3$. It is the coefficient of $Y$ on expanding \thetag8.

\head Quantisation\endhead

The Ponzano-Regge wavefunction is a function of the edge labels on the boundary
$\Sigma$ of $M$. The length of an edge is defined to be the value of the label
plus one half. Thus the edge lengths are discrete, taking the values
$j+{1\over2}$, where $j\in {1\over2}\Z$, $j\ge0$, indexes the representations
of $\SU(2)$.
These lengths determine a Euclidean signature metric for $\Sigma$. We consider
the equations satisfied when the edge lengths $l_1$, $l_2$, $l_3$ around a
vertex at which the three edges meet are varied, the other edges in $\Sigma$
having fixed lengths. The simplest example is for a single tetrahedron, when
$$\psi(l_1,l_2,l_3)=\left\{\matrix a&b&c\\l_1&l_2&l_3\endmatrix\right\},\tag9$$
the 6j-symbol, for fixed $a$,$b$,$c$.

\midinsert\centerline{\epsfbox{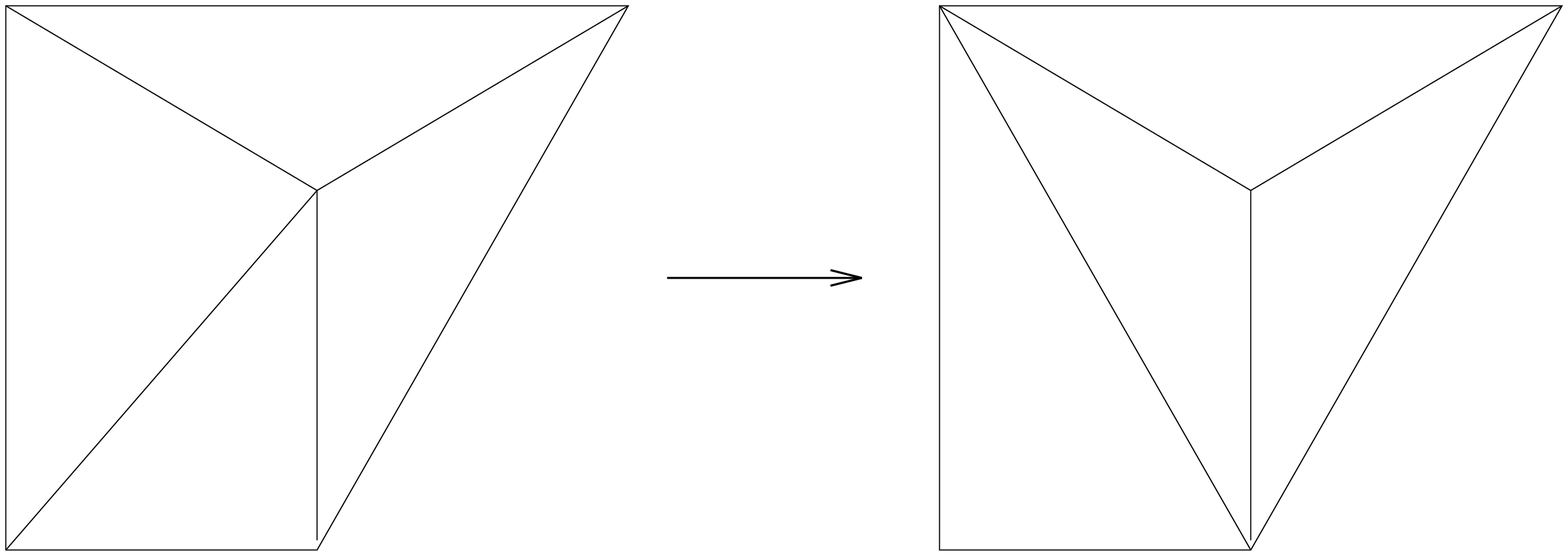}}\botcaption {Figure 3. Crossing
move}\endcaption \endinsert

Actually the case where three edges meet at a vertex suffices to handle the
general case, as the cases where more than three edges meet can be reduced to
this by the crossing move, shown in figure 3, which is an invertible linear map
$V\to V'$ and so generates no new equations \cite{3}. Also, any vertex at which
three edges meet reduces to the case \thetag9 because the Ponzano-Regge
wavefunction factorises as
$$\psi(l_1,l_2,l_3)=\left\{\matrix a&b&c\\l_1&l_2&l_3\endmatrix\right\}
\psi'(a,b,c,\text{other edges})$$
by the excision move, shown in figure 4 \cite3. Alternatively, this can be seen
as an expression of the invariance of the wavefunction under a change of
triangulation of the interior of $M$.

\midinsert\centerline{\epsfbox{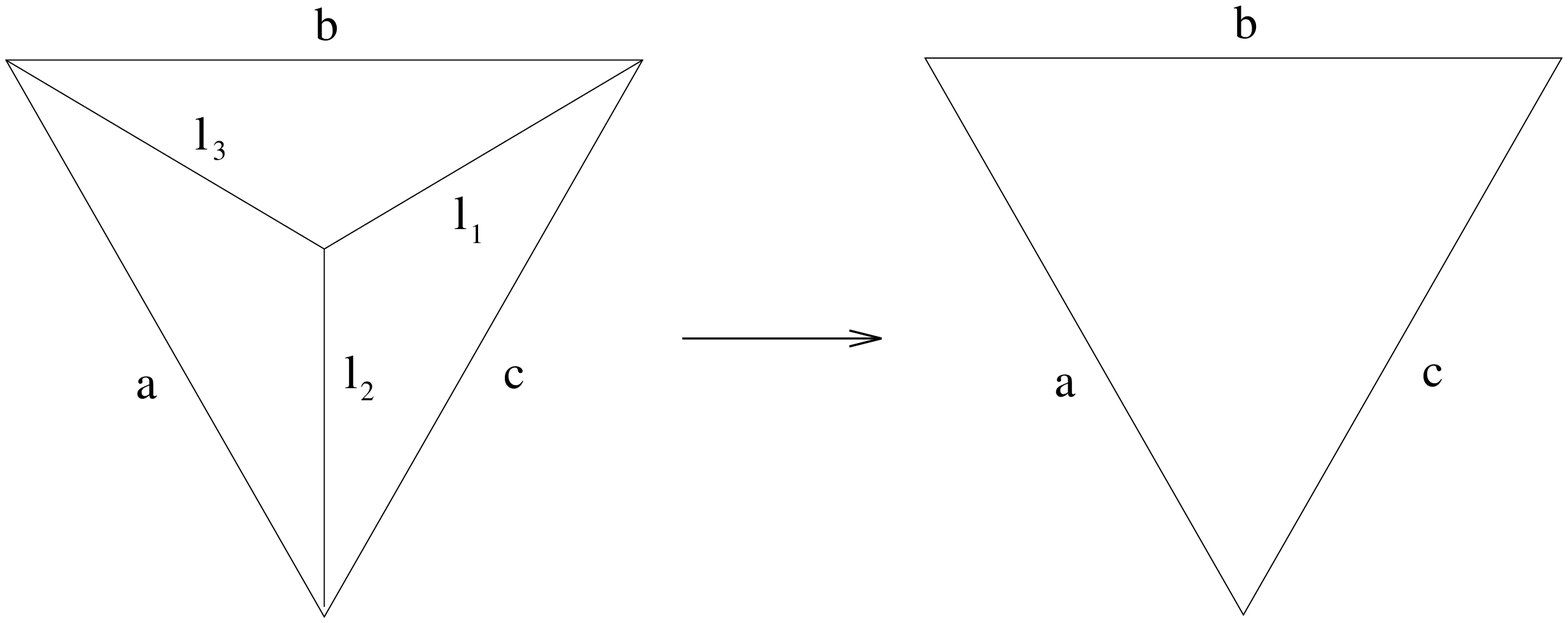}}\botcaption {Figure 4. Excision
move}\endcaption \endinsert

Our result for the Ponzano-Regge wavefunction is that for each representation
$J$ it satisfies an equation
$$D^{(23)}_{m_2m_3}\psi(l_1,l_2,l_3)=(-1)^{2J}\sum^J_{m_1=-J}
D^{(31)}_{m_3m_1}D^{(12)}_{m_1m_2}\psi(l_1+m_1,l_2+m_2,l_3+m_3)\tag{10}$$
In this formula, there are three square matrices of numerical coefficients
$D_{mn}$, of dimension $2J+1$. These matrices depend on the edge lengths
$(l_1,l_2,l_3,a,b,c)$. More specifically, $D^{(23)}$ depends on $(l_2,l_3,a)$,
$D^{(31)}$ depends on $(l_1,l_3,b)$, and $D^{(12)}$ depends on $(l_1,l_2,c)$,
so that each matrix is associated to one of the triangles in Figure 1.

\midinsert\centerline{\epsfbox{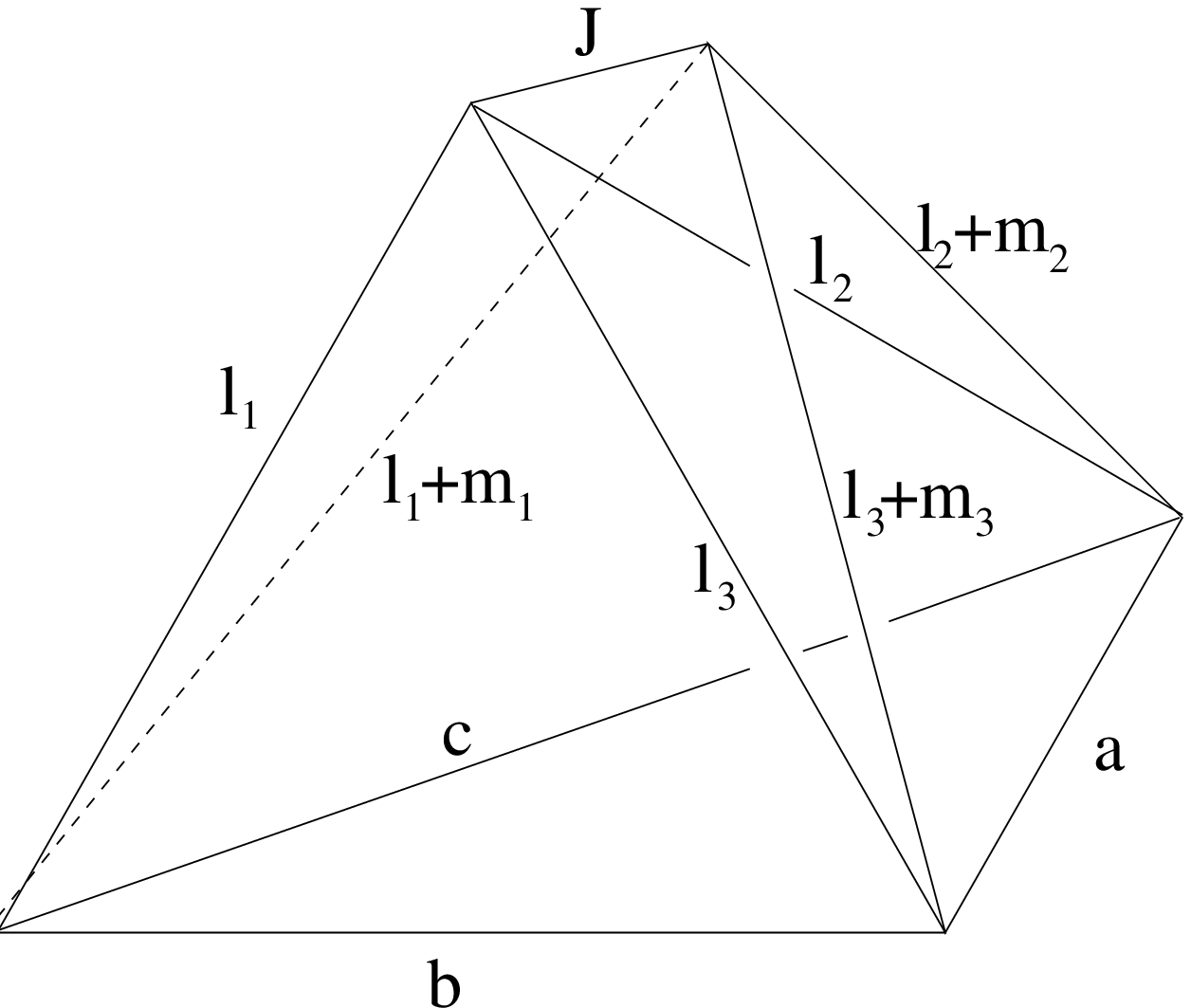}}\botcaption {Figure 5.
Complex}\endcaption \endinsert

\demo {Proof of formula \thetag{10}}
Apply the Biedenharn-Elliott relation to the complex shown in figure 5. This
gives
$$\multline\left\{\matrix a&l_3&l_2\\J&l_2+m_2&l_3+m_3\endmatrix\right\}
\left\{\matrix a&b&c\\l_1&l_2&l_3\endmatrix\right\}=
\sum^J_{m_1=-J}(-1)^\chi
(2l_1+2m_1+1)\\
\left\{\matrix b&l_1&l_3\\J&l_3+m_3&l_1+m_1\endmatrix\right\}
\left\{\matrix c&l_2&l_1\\J&l_1+m_1&l_2+m_2\endmatrix\right\}
\left\{\matrix a&b&c\\l_1+m_1&l_2+m_2&l_3+m_3\endmatrix\right\}\endmultline$$
with $\chi=a+b+c+l_1+l_2+l_3+J+l_1+m_1+l_2+m_2+l_3+m_3$.
This is equation \thetag{10}, with
$$D_{m_2m_3}^{(23)}=(-1)^{l_2+l_3+a+J+m_3}
(2l_2+2m_2+1)^{1\over2}(2l_3+2m_3+1)^{1\over2}
 \left\{\matrix a&l_3&l_2\\J&l_2+m_2&l_3+m_3\endmatrix\right\}$$
(and likewise for $D^{(31)}$, $D^{(12)}$).\enddemo

This equation can now be compared to \thetag{8}. We claim that it is a
quantisation of \thetag{8}. First, \thetag8 expressed in any irreducible
representation $J$. It can be written very explicitly using the weight basis in
which the $K$'s are diagonal.
$$K_n=\pmatrix
e^{iJ\phi_n}\\&e^{i(J-1)\phi_n}\\&&\ddots\\&&&e^{-iJ\phi_n}\endpmatrix$$
and the $R$'s then have the real matrix elements denoted $d^J_{mm'}(\theta)$ in
the angular momentum literature, related to the Jacobi polynomials. For
example,
$$d^{1\over2}(\theta)=\pmatrix
\cos\theta/2&\sin\theta/2\\-\sin\theta/2&\cos\theta/2\endpmatrix,$$
$$d^1(\theta)=\pmatrix
{1\over2}(1+\cos\theta)& {1\over\sqrt2}\sin\theta &{1\over2}(1-\cos\theta)\\
-{1\over\sqrt2}\sin\theta& \cos\theta& {1\over\sqrt2}\sin\theta\\
{1\over2}(1-\cos\theta)& -{1\over\sqrt2}\sin\theta &{1\over2}(1+\cos\theta)
\endpmatrix.$$
Equation \thetag8 becomes
$$d^J_{m_2m_3}(\theta_{23})=\sum^J_{m_1=-J}
e^{im_3\phi_3} d^J_{m_3m_1}(\theta_{31})
e^{im_1\phi_1} d^J_{m_1m_2}(\theta_{12})
e^{im_2\phi_2}.\tag{11}$$

According to the formula given by Edmonds \cite{15}, $D_{m_2m_3}^{(23)}$ is
asymptotic to $$d^J_{m_2m_3}(\theta_{23})=d^J_{m_3m_2}(-\theta_{23})$$
for fixed $m_2,m_3,J$ but $a,l_2,l_3\to\infty$. Similar formulae hold for
$D^{(12)}$, $D^{(31)}$. For this approximation to be valid, it is therefore
necessary to assume $J$ is much smaller than the six edge labels appearing in
Figure 1.

Then the  equation \thetag{10} is a quantisation of \thetag{11} in which
$e^{im\phi}$ is replaced by the shift operator $T$
$$T\psi(l)=\psi(l+m),$$
which was suggested as a heuristic by Ponzano and Regge.

The other noteworthy feature is the factor $(-1)^{2J}$. This is $-1$ for odd
spin representations and gives the extension of \thetag8 to $\SU(2)$ as the
equation
$$h^{\SU(2)}=-\identity.$$
This is the natural extension as the product of the matrices in \thetag8 gives
a rotation which is exactly one full turn, which lifts to $-\identity$ in
$\SU(2)$. This phenomenon is discussed in \cite{16}.

\subhead Remarks \endsubhead
Our quantisation of the discrete Wheeler-DeWitt equation gives an appealing
physical explanation of the Ponzano-Regge wavefunction by an analogy with other
quantised systems such as particle mechanics. But it is not clear exactly what
the mathematical `point' of our work is; can anything now be proved that was
not known before?

The connection between spin networks and 2+1
dimensional gravity has already been made in [1,2]. However this proceded in an
indirect manner from the Biedenharn-Elliott relation to form a
3-term recursion relation for 6j-symbols, and then to a semi-classical limit.
We have shown here that in the Hamiltonian
picture the analog of the Wheeler-DeWitt equation is exactly the
Biedenharn-Elliott relation itself. This has two advantages: firstly it makes
a more direct connection with the ideas of topological quantum field
theory, since it focuses on the consideration of a manifold with
boundary, and secondly it suggests an easier and more direct approach to
making physical interpretations of other topological state sums.
We note that the recursion relation occurs as a special case of our formula,
namely when $m_2=m_3=0$ and $J=1$.

Our quantisation appears to give more information than the semiclassical
formula for the 6j symbols given by Ponzano and Regge. For example, it gives an
explanation for the discreteness of the edge length labels, which is not
required by the semiclassical formula. Normally one would quantise the variable
conjugate to the length $l$ by $\phi\to \d/\d l$. Then $l$ would have to have a
continuous spectrum. However, we are interested in quantising $\exp(im\phi)$,
which, making the above substitution, gives the shift operator, given by the
action $l\to l+m$. As $m$ is a half-integer, this operator natually acts on
functions defined only on the discrete set of points ${1\over2}\Z$.

Although we have considered quantising the formulae obtained from
3-di\-men\-sion\-al metrics with a Euclidean signature, one can also start with
the corresponding Lorentzian formulae. Then there are some sign differences in
\thetag3 to \thetag5, which are reflected in the fact that \thetag8 becomes an
equation in $\SO(2,1)$ with the $K$'s boosts and the $R$'s unchanged, c.f.
\cite{17}. The quantisation replaces the matrix element $e^{J\phi}$ with a
shift operator, and one obtains the same asymptotic approximation to
\thetag{10}.

\enddocument